\pdfoutput=1
\PassOptionsToPackage{hyphens}{url}
\documentclass{article}
\usepackage{arxiv}
\usepackage[utf8]{inputenc}
\usepackage[T1]{fontenc}
\usepackage{hyperref}
\usepackage{url}
\usepackage{graphicx}
\usepackage{xcolor}
\usepackage{amsmath,amssymb,amsfonts}
\usepackage{booktabs,tabularx}
\usepackage{microtype}
\usepackage[square,numbers,sort&compress]{natbib}
\definecolor{boxhead}{HTML}{333333}
\definecolor{boxbg}{HTML}{FAF9F6}
\newcounter{box}
\newcounter{mainproposition}
\newcounter{supptheorem}
\newcounter{suppsection}
\renewcommand{\thesuppsection}{S\arabic{suppsection}}
\newcounter{supptable}

\title{The Ineliminable Anchor: Human Judgment as a Structural Requirement of Autonomous Artificial Intelligence Systems}
\date{}

\usepackage{authblk}

\author[1]{Junjie Li}
\author[1]{Haoran Liu}
\author[1]{Yidan Zhang}
\author[2]{Jian Xu}
\author[1,3]{Zhaoyuan Ma\thanks{Corresponding author: \texttt{mazy@sustech.edu.cn}}}
\affil[1]{School of Automation and Intelligent Manufacturing, Southern University of Science and Technology, Shenzhen, China}
\affil[2]{School of Design, Southern University of Science and Technology, Shenzhen, China}
\affil[3]{University College, Korea University, Seoul, Korea}

\renewcommand{\shorttitle}{The Ineliminable Anchor}
\hypersetup{
pdftitle={The Ineliminable Anchor: Human Judgment as a Structural Requirement of Autonomous Artificial Intelligence Systems},
pdfauthor={Junjie Li, Haoran Liu, Yidan Zhang, Jian Xu, Zhaoyuan Ma},
hidelinks,
}

\begin{document}
\maketitle

\begin{abstract}
As artificial intelligence (AI) agents carry out longer chains of tasks, oversight must establish whether their behavior still serves an authorized goal. Here, we introduce the anchor: the point at which reference information and human-authorized standards enter an autonomous system. We argue that oversight design should first identify the reference needed to assess goal drift and whether the system's internal record can supply it, then decide who checks and how often, and finally specify who may revise the goal. Goal drift can be impossible to certify from a system's internal record. The missing reference must come from outside that record. Combining a reliability model of checkpoint spacing with evidence from training, execution, and multi-agent coordination, we develop a task taxonomy to guide anchor placement and decisions about which checks to automate. For AI systems that serve human purposes, human judgment is an ineliminable structural requirement in authorizing and revising the purposes and standards that govern autonomous operation, even when routine checking and execution are automated.
\end{abstract}

\section{Introduction}\label{sec:introduction}

Progress in artificial intelligence (AI) is often framed as progress in capability: what tasks machines can do and how well they perform them. We focus on a different question: do a system's actions remain faithful to its stipulated goal? Answering it means comparing the system's behavior with something trusted, such as an authenticated target identity, reliable labels, or an authorized specification. We call this reference information. We argue that oversight design should first establish what reference information a judgment needs and whether the system's own record can supply it, then decide who checks and how often, and finally specify who may revise the goal.

We consider AI systems that serve human purposes. Beneficiaries are those whose interests the system should serve, and the principal is the person or institution authorized to establish or revise the system's goal standard, possibly on their behalf. Instructions, intentions, revealed and ideal preferences, interests, and values are distinct alignment targets~\cite{gabriel2020}. We adopt the principal's stipulated standard as the goal $G$ that the system pursues. Even when the application of that standard is delegated, its authority remains traceable to people or institutions, which gives human judgment a structural role.

Article 14 of the EU AI Act requires high-risk AI systems to be designed for effective oversight by natural persons during use, with oversight measures proportionate to the risks, level of autonomy, and context of use~\cite{eu2024}. Bainbridge's irony of automation shows why this is hard: automation reduces the opportunities to practice while leaving operators the hardest exceptions~\cite{bainbridge1983}. A similar challenge can arise with autonomous AI\@. Designers of AI systems, therefore, need to identify which of a system's judgments require additional reference information and give the people who make them evidence, practice, and effective methods of intervention.

AI has greatly increased the speed and volume of output, and with it the demand for review. Google reports that 75\% of its new code is now generated by AI and approved by engineers (\url{https://blog.google/innovation-and-ai/infrastructure-and-cloud/google-cloud/cloud-next-2026-sundar-pichai/}). Whether verification works depends on reference information. A Deloitte report published by the Australian government contained a fabricated quotation from a federal court judgment, which came to light only after a university researcher alerted the media~\cite{mcguirk2025deloitte}, and early in 2026 about one in 277 biomedical papers cited a work that does not exist~\cite{topaz2026}. Such errors are almost impossible to recognize within the document and show only when the cited judgment or paper itself is checked. Checks also have to come at the right time. In METR's 2025 measurements, frontier AI agents completed nearly all tasks that take people under four minutes but fewer than 10\% of those taking more than about four hours~\cite{metr2025blog}, so checks on long tasks have to be spread along the run and placed before irreversible actions. Otherwise the consequences can be severe, as when Replit's coding agent deleted a production database without authorization during a code freeze~\cite{nolan2025replit}. Finally, authority and responsibility for the system have to rest with people or institutions. Air Canada's chatbot told a passenger that a discounted fare could be claimed after the trip, which the airline did not in fact allow. When the airline argued that the chatbot was responsible for its own statements, the tribunal rejected the argument~\cite{moffatt2024}. An institution answers for any standard its system states and must specify who may set and revise those standards. Verification is already falling behind output, so allocating limited review capacity properly has become all the more important.

All three problems arise where reference information and judgment enter the system. We call such a point an anchor. An anchor can shift from runtime to design time, from an operator to a verifier, or into an institutional procedure. Data, sensors, verifiers, and procedures can carry and apply the principal's standards, so checking and execution can be automated once validated. Shneiderman's human-centered AI framework likewise treats automation and human control as independent dimensions that can both be high~\cite{shneiderman2020}.

The following sections pursue this design sequence. Starting from Turing's ordinal logics, we examine the limits on certifying that a system keeps its goal. We then use the anchor model to trace where errors compound in training, execution, and coordination, and from this derive how oversight should be matched to the task.

\section{What Rules Cannot Justify}\label{sec:forgotten-1939}

\begin{figure*}[tp]
\refstepcounter{figure}\label{fig:ordinal-logics}
\centering
\includegraphics[width=\textwidth]{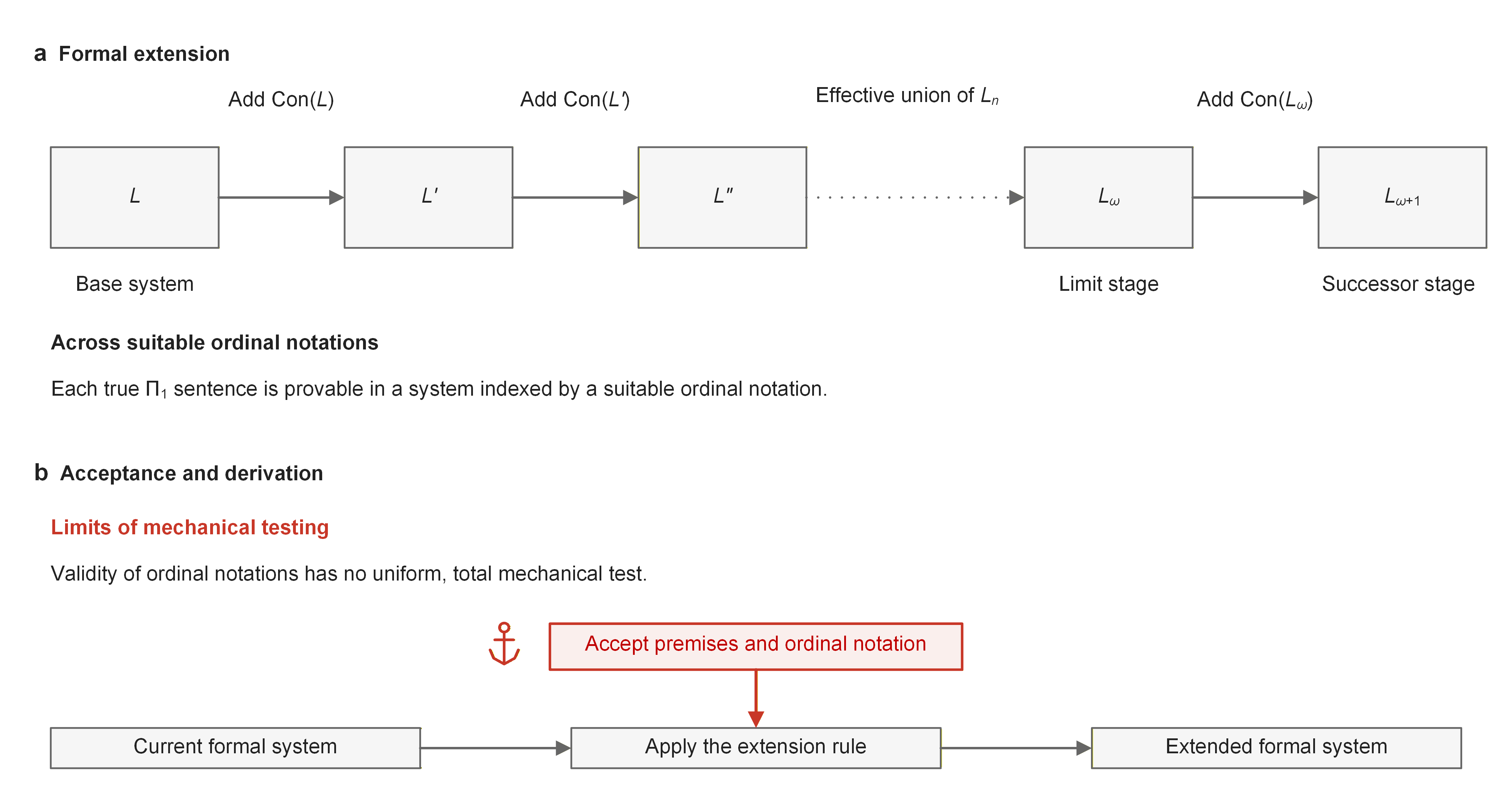}
\par\smallskip\begin{minipage}{\textwidth}\footnotesize\raggedright
\textbf{Fig. \thefigure{} \textbar{}} Turing's ordinal logics: formal extension and accepted premises. a, Each successor stage adds the consistency statement $\mathrm{Con}(L)$ of the current system, and the limit stage $L_\omega$ is the effective union of the preceding finite stages $L_n$. The dotted arrow abbreviates intermediate stages. Completeness holds across suitable ordinal notations, not along every path~\cite{turing1939,feferman2006}. b, The extension rule takes accepted premises and an ordinal notation as inputs. The red anchor marks where these inputs are accepted. Once they are fixed, formal derivation follows the rules.
\end{minipage}
\end{figure*}

Mathematical logic gave an early and precise account of why a system that follows rules needs input from outside itself. G\"odel's incompleteness theorems show that any consistent formal system that is strong enough for arithmetic, and whose rules can be listed mechanically, leaves some true statements unprovable within it, among them the statement that the system is consistent~\cite{godel1931}. Turing's ordinal logics respond by adding that statement as a new axiom, which gives a stronger system, and then repeating the step (Fig.~\ref{fig:ordinal-logics}). If each step is chosen well, every true statement of a basic kind, such as one asserting that a checkable property holds for all natural numbers, becomes provable at some step~\cite{turing1939,feferman2006}. Feferman later showed that iterating a stronger principle, uniform reflection, extends this to every true statement of arithmetic~\cite{feferman1962,feferman2006}. Adding the axiom is mechanical, but deciding how to continue, namely which of the ordinal notations that label the steps are valid, is something no mechanical test, and hence no Turing machine, can always get right~\cite{feferman2006}.

Turing called the capacity that rule-governed operations require \emph{ingenuity}, and the capacity needed to accept an extension, especially to recognize a valid notation, \emph{intuition}~\cite{turing1939}. He also described an oracle machine: an ordinary machine with an interface through which it can ask questions it cannot decide by itself.

Oversight of AI systems faces the same divide. Today's AI models, and the monitors that check them, run on digital computers. Anything they compute can also be computed by a Turing machine~\cite{turing1936,hopcroft2007}, so they are mechanical in exactly this sense. A monitor that reads a system's record applies rules, and in some cases the record cannot settle whether the system has drifted from its goal. Those cases need reference information from outside the record, supplied at an anchor.

\begin{table*}[tp]
\refstepcounter{table}\label{tab:frameworks}
\centering
\begin{minipage}{\textwidth}\footnotesize\raggedright
\textbf{Table \thetable{} \textbar{}} The anchor and neighboring frameworks.
\par\medskip
\begingroup
\setlength{\tabcolsep}{4pt}
\renewcommand{\arraystretch}{1.08}
\begin{tabularx}{\textwidth}{@{}>{\raggedright\arraybackslash}p{0.19\textwidth}>{\raggedright\arraybackslash}X>{\raggedright\arraybackslash}X@{}}
\toprule
\textbf{Framework} & \textbf{Scope} & \textbf{Reference and design question} \\
\midrule
Human-in/on-the-loop~\cite{chiodo2026,bainbridge1983} & Roles, workload, and interfaces for human involvement & Which judgments need additional references, who can supply them, and what checking schedule do the reliability assumptions support? \\
\addlinespace[5pt]
Formal verification \& runtime monitoring~\cite{clarke1986,seshia2022,dalrymple2024} & Guarantees relative to models and specifications; design-time and runtime assurance & What evidence connects the specification and environment model to the principal's standard and deployment conditions? \\
\addlinespace[5pt]
Feedback control \& observability~\cite{luenberger1964} & State estimation under model and observability assumptions & Can admissible cases share the observable record yet require different goal-relative drift verdicts (Proposition~\ref{prop:internal-certification})? \\
\addlinespace[5pt]
Goal alignment \& reward learning~\cite{hadfield2016} & Objective inference from human behavior and interaction under reward uncertainty & What evidence connects the learned objective to an authorized standard and beneficiaries' interests? \\
\addlinespace[5pt]
Responsibility attribution~\cite{sterz2024,eu2024} & Legal and ethical accountability; effective oversight & Does the responsible party have the evidence, review, override, and abort authority needed to exercise its role? \\
\bottomrule
\end{tabularx}
\endgroup
\end{minipage}
\end{table*}

Table~\ref{tab:frameworks} places the anchor among five neighboring frameworks, two of which are closest to it. Chiodo and colleagues use oracle machines to model what human input contributes to a system, its failure modes, and legal-moral responsibility in human-in-the-loop oversight~\cite{chiodo2026}. Formal verification starts from a specification~\cite{clarke1986,seshia2022,dalrymple2024}. Both approaches take their input as given: the human oracle in one case, the specification in the other. But when is such an input needed? What justifies relying on it? These are the questions the anchor framework addresses: whether a record settles a run's verdict, what information would, whether the specification matches the authorized goal, and who may revise that goal.

Whether it is a formal theory or a monitor, a system that follows rules cannot make every judgment it needs from its own rules and record alone. What is missing has to come from outside. 

\section{The Anchor Model}\label{sec:anchor-model}

\begin{figure*}[tp]
\refstepcounter{figure}\label{fig:anchor-model}
\centering
\includegraphics[width=\textwidth]{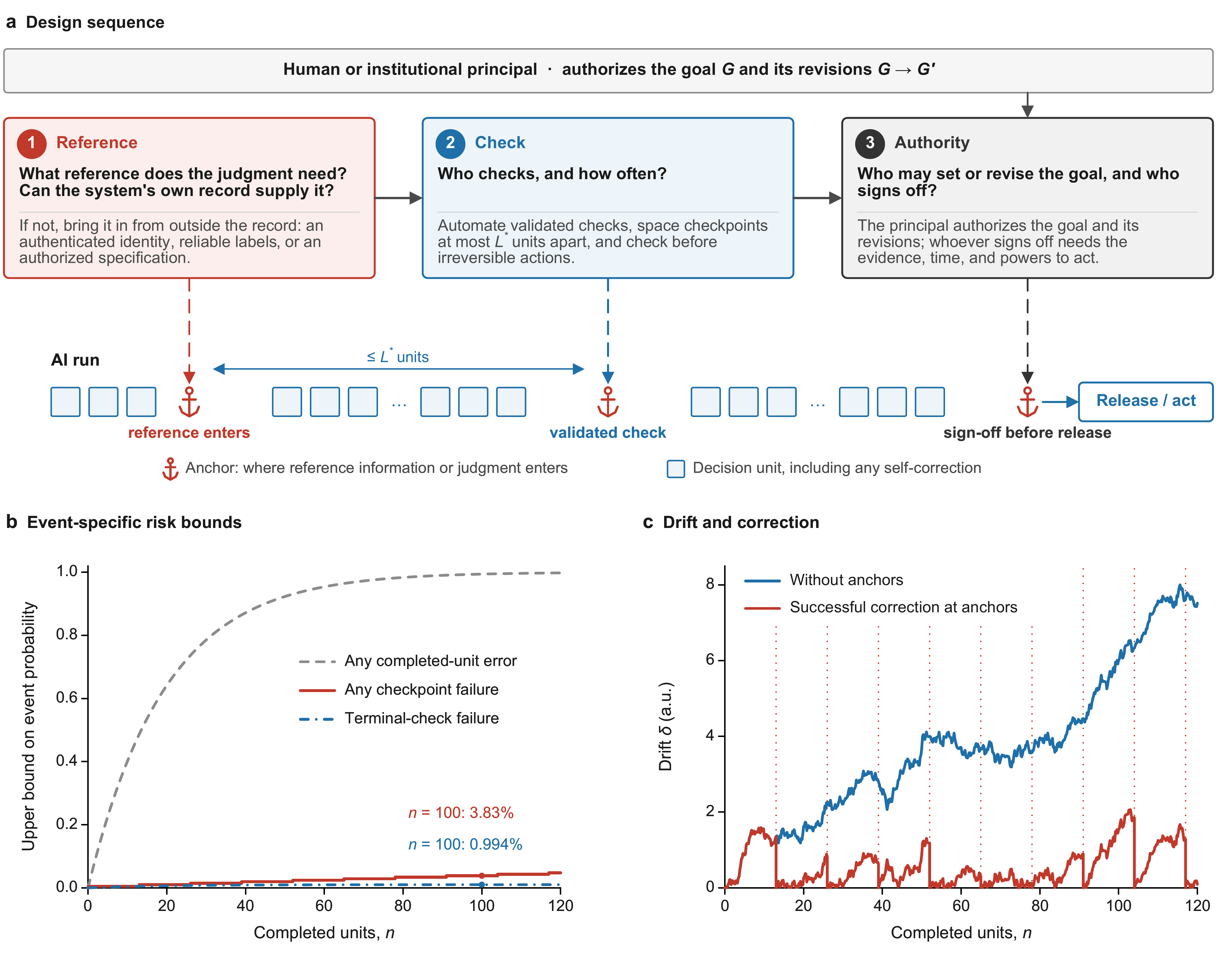}
\par\smallskip\begin{minipage}{\textwidth}\footnotesize\raggedright
\textbf{Fig. \thefigure{} \textbar{}} Anchors and segmented design. a, Design sequence: the three questions a designer answers in order, and where each places an anchor in the run. b, Model upper bounds on three failure events over $n$ completed units: at least one unit error, including errors later repaired (gray); at least one checkpoint left incorrect after correction, with checkpoints $L=L^*=13$ units apart (red); and failure after a single terminal check (blue). Parameters are illustrative: $p=0.95$, $q=0.99$, and $\tau=0.5$. Box~\ref{box:formal-results} and Supplementary Section~\ref{sec:si-segment-length} give the bounds and their assumptions. c, Schematic drift without anchors (blue) and with successful correction at each anchor (red). Dotted lines mark the anchors.
\end{minipage}
\end{figure*}

\begin{figure*}[tp]
\refstepcounter{box}\label{box:formal-results}
\centering
\fcolorbox{boxbg}{boxbg}{\begin{minipage}{\dimexpr\textwidth-2\fboxsep-2\fboxrule\relax}
\colorbox{boxhead}{\makebox[\dimexpr\linewidth-2\fboxsep\relax][l]{\rule[-4pt]{0pt}{15pt}\textcolor{white}{\sffamily\bfseries BOX \thebox\quad{\normalfont\textbar}\quad The formal results behind anchor design}}}
\par\medskip\footnotesize
\begin{minipage}[t]{0.485\linewidth}
\par\addvspace{6pt}\refstepcounter{mainproposition}\label{prop:internal-certification}
\noindent\textbf{Proposition \themainproposition{} (limits of internal drift certification).} If two admissible runs share the complete internal record but differ on whether drift is absent (A1), a monitor reading only that record (A2) cannot provide universally correct binary certification. Identical inputs require the same output; adding monitors that read the same record preserves this ambiguity. Resolving it requires additional reference information that separates the cases. Supplementary Section~\ref{sec:si-indistinguishability} gives the proof and the scope for statistical monitoring.
\par\smallskip
\noindent\textbf{Prospective certification.} Supplementary Theorem~\ref{thm:prospective-certification} uses a halting reduction~\cite{turing1936}: over a class containing the constructed simulator family, no algorithm terminates on every program description and correctly decides whether a fixed goal is preserved throughout the future. Combining computable verifiers does not overcome this limit.
\par\smallskip
\noindent\textbf{Checkpoint spacing.} We use decision units of fixed granularity: a unit may include execution, checking, and internal self-correction, and its correctness criterion must cover relevant irreversible consequences. If each completed unit has a conditional correctness probability of at least $p$, given that all preceding units were correct, the probability that all $n$ units are correct is at least $p^n$. If correctness of every unit suffices for task success, recovery across units can further increase final success. Supplementary Section~\ref{sec:si-segment-length} uses these distinctions to define segment guarantees and cumulative checkpoint risk.
\end{minipage}\hfill
\begin{minipage}[t]{0.485\linewidth}
Now place anchors at chosen points, dividing the chain into segments. For a segment starting in a correct state, assume that every completed unit has a uniform conditional correctness lower bound $p$, given correct preceding units. The probability that all units in a segment of length $L$ are correct is then at least $p^L$, without an independence assumption. We adopt the design criterion that this guarantee, before external correction, meets a chosen target $\tau$ for each segment. The largest length supported by this criterion is
\begin{equation}\label{eq:segment-length}
L^*=\left\lfloor\frac{\ln\tau}{\ln p}\right\rfloor,
\qquad 0<p<1,\quad 0<\tau<1.
\end{equation}
\noindent The same logarithmic relation appears in Sinha et al.'s analysis of task horizon~\cite{sinha2026}. The floor gives the largest integer $L$ with $p^L\ge\tau$, so $L^*$ depends on the unit definition, the validated $p$, and the chosen $\tau$; at $p=0.95$ and $\tau=0.5$, $L^*=13$.
\par\smallskip
\noindent The per-segment target and overall task reliability concern different events. If, given earlier checkpoint successes, each anchor corrects a still-incorrect input with conditional probability at least $q$ and preserves correct inputs, the cumulative risk $R_k$ that any of the $k=\lceil n/L\rceil$ checkpoints fails satisfies $R_k\le1-[p^L+(1-p^L)q]^k$ (Supplementary Section~\ref{sec:si-segment-length}; Fig.~\ref{fig:anchor-model}b). For the illustrative 100-unit task, $k=8$ and $R_8\le0.0383$.
\end{minipage}
\end{minipage}}
\end{figure*}

This section addresses three questions on which the first two steps of the design sequence (Fig.~\ref{fig:anchor-model}a) depend. When does an internal record fail to settle whether the goal has drifted? Can it be decided uniformly whether a system will keep its goal throughout an unbounded future? How can a reliability target bound the spacing between checkpoints? The first question determines when an outside reference is needed, and the other two determine why checks have to take place during operation and how often. Who may revise the goal is a normative question, taken up in Section~\ref{sec:oversight-taxonomy}.

The anchor model has four parts (Fig.~\ref{fig:anchor-model}). First, a system acts in decision units, each of which may include execution, checking, and self-correction within the unit, and it leaves an internal record: every input, state, output, and log it can read. Second, the principal stipulates a success standard $G$. Drift occurs when the goal represented in the record, read in the system's environment, departs from $G$. Third, an internal monitor reads only that record, and certification requires it to be correct on every admissible run, that is, every run allowed by the stated deployment assumptions and by what the monitor can access. Fourth, an anchor is a point where reference information or judgment from outside the record enters. It can supply what the record lacks and can detect and correct errors. The units between two consecutive anchors form a segment. In this model the three questions become: whether the record determines the drift verdict; whether an algorithm can decide from a system's description that drift never occurs; and, given unit reliability and a target for each segment, how many units a segment may contain. Box~\ref{box:formal-results} states the three results, and Supplementary Section~\ref{sec:si-definitions} gives the formal definitions.

The record fails to settle the verdict when two admissible runs leave the same record but differ on whether drift has occurred: a monitor that reads only the record cannot certify both correctly, and only information from outside the record can separate them (Proposition~\ref{prop:internal-certification} in Box~\ref{box:formal-results}). Take a reference in a paper's bibliography. Whether the cited paper exists is a fact about the published literature, not about the citing paper. The same complete-looking entry, with authors, title, journal, and year, could point to a real paper or to one that does not exist. A checker who reads only the citing paper sees the same text in both cases and cannot tell which is true. Only a search of the literature can settle it. Supplementary Section~\ref{sec:si-definitions} gives a formal construction of such a pair. The outside reference needed for that search is what an anchor brings in.

Other fields apply the same remedy. Under explicit assumptions, letting a system interact with people, or inferring the designer's intent from the reward they specified, improves estimates of the goal~\cite{hadfield2016,hadfield2017ird}. In constructed test tasks, giving a monitor replayable evidence and an explicit comparison rule improved monitoring~\cite{xu2026oversight}. Each of these brings in information from outside the record, which is what this Perspective calls an anchor.

Whether a system will preserve its goal throughout an unbounded future cannot be decided uniformly. Take any class of systems that contains, for every program, a system that simulates it and switches goals if it halts. No algorithm can decide for every system in that class whether a fixed goal will always be preserved, because such an algorithm would also decide the halting problem, which no Turing machine can do~\cite{turing1936} (Box~\ref{box:formal-results}). Supplementary Theorem~\ref{thm:prospective-certification} gives the proof, and Supplementary Remark R1 relates this limit to Rice's theorem~\cite{rice1953}. Because complete certification cannot be done in advance, oversight has to check during operation, which leaves the question of how far apart the checks may be.

The spacing follows from a reliability target for each segment. If each decision unit, such as one edit--test--retry cycle of a coding agent, is correct with conditional probability at least $p$ given correct preceding units, a segment of $L$ units that starts from a correct state is entirely correct with probability at least $p^L$. If this guarantee must reach a target $\tau$, set by the task's risk tolerance, before any external correction, the largest admissible spacing is $L^*$ (equation~\eqref{eq:segment-length} in Box~\ref{box:formal-results}). With $p=0.95$ and $\tau=0.5$, for example, $L^*=13$, and Fig.~\ref{fig:anchor-model}c illustrates drift accumulating between anchors spaced this far apart and falling back after each successful correction. Box~\ref{box:formal-results} also bounds the cumulative risk across checkpoints, which still grows with task length (Fig.~\ref{fig:anchor-model}b).

\section{Where Errors Compound and Anchors Belong}\label{sec:closed-loops}

Model training and the execution and coordination of AI agents are three concrete settings for anchor design. This section shows how errors compound in each and where anchors belong (Fig.~\ref{fig:closed-loops}). The three settings pose one design problem: wherever errors can compound, the system needs a point at which information tied to the intended goal re-enters it.

\subsection{Training}
When AI-generated data replaces original data across generations, models can collapse over successive iterations~\cite{shumailov2024}. In recursive fine-tuning of the OPT-125m language model, retaining 10\% original data coincided with lower perplexity (Fig.~\ref{fig:closed-loops}a). Whether collapse occurs in this training mode depends on which real data remain and how closely the synthetic data match them. Self-consuming training can lose quality or diversity without sufficient fresh real data each generation~\cite{alemohammad2024}. Test error can stay bounded in models trained on all accumulated real and synthetic data, but fixed-size subsets can still degrade~\cite{gerstgrasser2024,kazdan2025}. A mismatch between real and synthetic data distributions can impose a non-vanishing test-error floor in the regression models studied~\cite{dohmatob2025}. For training generative AI models, anchor design therefore has to decide which original or authenticated data to retain so that the training signal stays tied to the goal that people have set or authorized.

Validation data with the wrong coverage can also leave deployment behavior underdetermined. Predictors that perform similarly on validation data drawn from the training distribution can behave differently under stress tests that simulate deployment~\cite{damour2022}. In goal-misgeneralization experiments, policies kept their navigation skills after the environment changed yet pursued correlates of the training reward instead of the objective their designers intended~\cite{langosco2022}. Both findings show that the coverage of validation data needs thorough checking.

\subsection{Execution}
METR measures how AI agents perform on long tasks against the time skilled humans need to complete them: the 50\% horizon is the task length at which an agent succeeds with 50\% probability, counting the agent's own recovery during a run~\cite{kwa2025}. In March 2026, METR reported 50\% and 80\% horizons of approximately 55 and 20 minutes, respectively, for Claude 3.7 Sonnet on its time-horizon-1.0 suite~\cite{metr2026horizons} (Fig.~\ref{fig:closed-loops}b). Calibrated to the decision units of a specific task, such horizons can inform estimates of $p$ and $L^*$.

Errors can also reinforce themselves within a run: in experiments that supplied task knowledge and plans, earlier errors raised later per-step error rates in non-thinking models, whereas Qwen3's accuracy at turn 100 in thinking mode stayed relatively stable~\cite{sinha2026}. Whether self-correction helps depends on where the feedback comes from and on how the model was trained. Prompting tested models to reconsider their answers can spoil correct answers, and using ground-truth labels to decide when to stop changes the evaluation problem itself~\cite{huang2024}. Evaluations of a system and its self-correction should therefore record both repaired errors and spoiled correct answers, and distinguish training feedback from information available during execution.

\subsection{Coordination}
\begin{figure*}[tp]
\refstepcounter{figure}\label{fig:closed-loops}
\centering
\includegraphics[width=\textwidth]{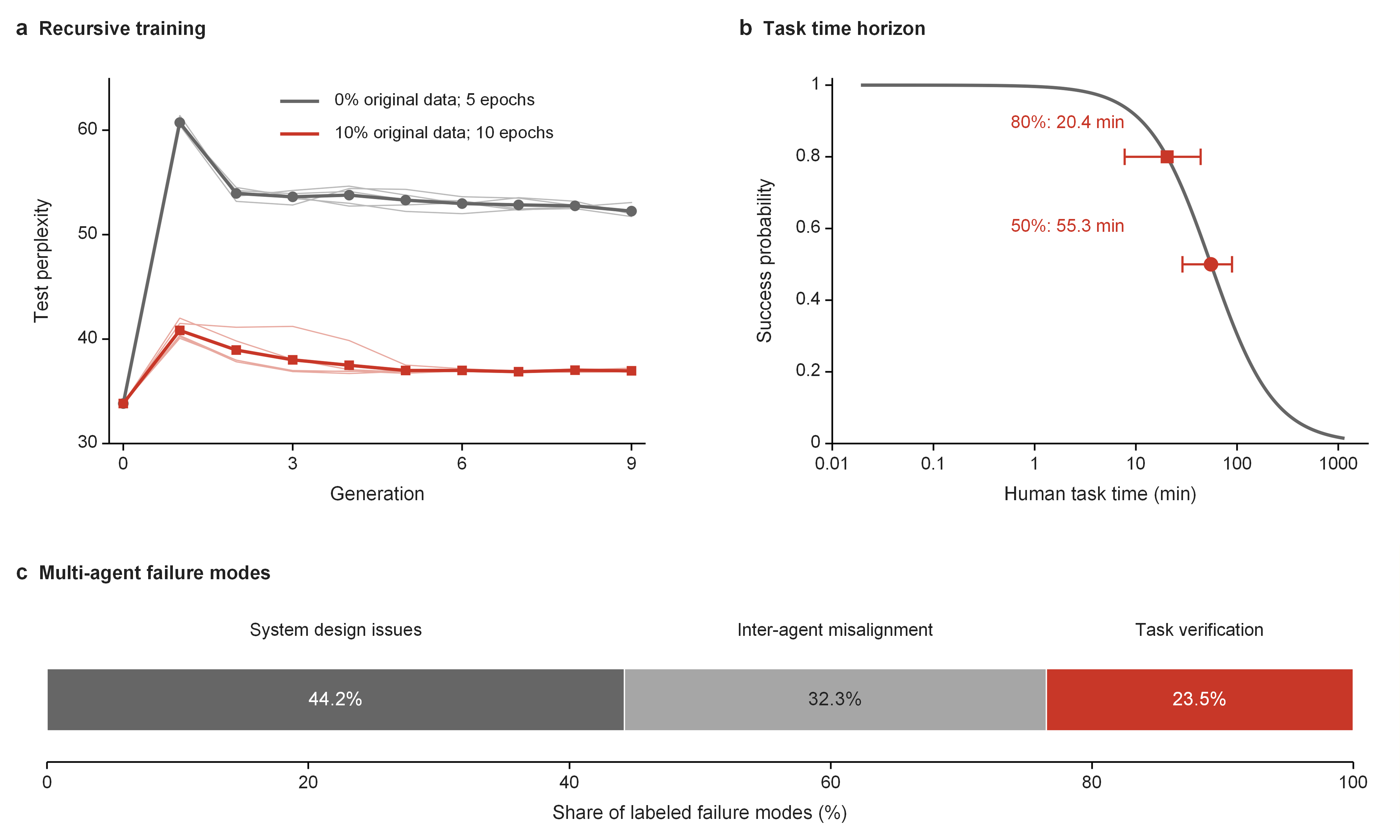}
\par\smallskip\begin{minipage}{\textwidth}\footnotesize\raggedright
\textbf{Fig. \thefigure{} \textbar{}} Evidence from training, execution, and coordination. a, Training: recursive fine-tuning of OPT-125m on WikiText-2~\cite{shumailov2024}, retaining 0\% of the original data with five epochs per generation (gray) or 10\% with ten epochs (red), so retention and training duration both differ. Thin lines are five run means per protocol and thick lines their unweighted mean. Generation 0 is the original-data baseline. Adapted from ref.~\cite{shumailov2024} under CC BY 4.0, with colors and aggregation modified. b, Execution: METR's corrected logistic fit for Claude 3.7 Sonnet on the 169-task time-horizon-1.0 suite (1,353 runs)~\cite{kwa2025}, reproduced from METR's public run data (Data Availability). Points mark the 80\% and 50\% horizons, and bars are METR's 95\% bootstrap confidence intervals~\cite{metr2026horizons}. c, Coordination: failure-mode shares among MAST's 1,642 LLM-annotated traces~\cite{cemri2025}. The LLM annotations agree with human annotators at $\kappa=0.77$.
\end{minipage}
\end{figure*}

In multi-agent systems, errors cross agent boundaries: one agent's output becomes another's input, and an error that is never checked in any form can travel down the chain. The Multi-Agent System Failure Taxonomy (MAST) organizes 1,642 traces from seven frameworks into 14 failure modes~\cite{cemri2025}. Across all traces labeled by a large language model (LLM), the failures fall into system design issues, inter-agent misalignment, and task verification (Fig.~\ref{fig:closed-loops}c). These categories correspond to three places where anchors can sit: the task and role specifications given to each agent, the handoffs between agents, and the completion criteria against which the joint output is checked.

Interactions among components can also produce failures that no single component shows. Hidden feedback loops connect local machine-learning decisions to wider system behavior~\cite{sculley2015}, deployed predictions shift the data distributions that follow~\cite{perdomo2020}, and accidents can arise from interactions among controllers even when every component works as designed~\cite{leveson2004}. Checking each component therefore cannot replace checking at the system level: anchors at component boundaries have to bring feedback about those consequences to someone who can act on it.

\section{Matching Oversight to the Task}\label{sec:oversight-taxonomy}

\begin{table*}[tp]
\refstepcounter{table}\label{tab:taxonomy}
\centering
\begin{minipage}{\textwidth}\footnotesize\raggedright
\textbf{Table \thetable{} \textbar{}} A task taxonomy for anchor placement. The dimensions are output verifiability $V_o$, chain length and irreversibility $R_c$, goal contestability $C_g$, and accountability weight $A_w$, distinct from the goal $G$ and the risk $R_k$. Profiles are illustrative combinations, and tiers denote oversight needs that can coexist. Levels are high (H), medium (M), or low (L)\@. $V_o$: per-output checks covering the stated correctness criterion at the required error level (H), sampled checks or partial coverage of that criterion (M), or predominantly contextual judgment (L)\@. $R_c$: long chains or major irreversible actions (H), error propagation with effective recovery (M), or short and readily recoverable execution (L)\@. $C_g$: substantive contestation (H), remaining interpretive choices (M), or settled objectives (L)\@. $A_w$: high-consequence decisions, or decisions requiring formal legal or regulatory sign-off, with a designated accountable authority (H; designation alone is insufficient); consequential outcomes requiring organizational review or audit (M); or minor external consequences with routine accountability (L)\@. n/a: the goal to be verified is still being stipulated, although goal proposals can be evaluated.
\par\smallskip
\begingroup
\setlength{\tabcolsep}{4pt}
\renewcommand{\arraystretch}{1.08}
\begin{tabularx}{\textwidth}{@{}>{\raggedright\arraybackslash}p{0.13\textwidth}>{\raggedright\arraybackslash}p{0.10\textwidth}>{\raggedright\arraybackslash}X>{\raggedright\arraybackslash}X>{\raggedright\arraybackslash}X@{}}
\toprule
\textbf{Tier} & \textbf{Illustrative profile} & \textbf{Representative tasks} & \textbf{Anchor placement and mode} & \textbf{Issue addressed} \\
\midrule
1. Mechanically verifiable & $V_o$: H\newline $R_c$: L\newline $C_g$: L\newline $A_w$: L--M & Proof checking; compilation and formal verification; settlement reconciliation & Design-time specification and verifier review; runtime anchor density can approach zero within verified scope & Specification fidelity to intent; system-design failures~\cite{cemri2025} \\
\addlinespace[5pt]
2. Statistically auditable & $V_o$: M\newline $R_c$: M\newline $C_g$: L--M\newline $A_w$: M & Recommendation; classification; risk pre-screening; content triage & Periodic sampled audits and continuous monitoring; audit design sets coverage and frequency & Potentially unidentifiable drift (Proposition~\ref{prop:internal-certification}); distribution shift; information loss in recursive training~\cite{shumailov2024} \\
\addlinespace[5pt]
3. Long-chain open-ended & $V_o$: L--M\newline $R_c$: H\newline $C_g$: M\newline $A_w$: M--H & Long-horizon agents; autonomous driving; automated operations & Spacing supported by $L^*$ where assumptions hold; checks before irreversible actions; automatic safeguards and human exception handling & Time horizons~\cite{kwa2025}; error self-conditioning in non-thinking models, mitigated in thinking experiments~\cite{sinha2026} \\
\addlinespace[5pt]
4. Goal-generating & $V_o$: n/a\newline $R_c$: M\newline $C_g$: H\newline $A_w$: H & Research agendas; product definition; policy goals; curriculum design & Principal retains authority to establish and revise goals; AI supports analysis and advice & Contested objectives and authorization; performative feedback~\cite{perdomo2020} \\
\addlinespace[5pt]
5. Accountability-terminal & $V_o$: M\newline $R_c$: H\newline $C_g$: M--H\newline $A_w$: H & Medical sign-off; judicial decisions; safety-critical engineering release & Final human sign-off where governing rules require an answerable person & Responsibility allocation, automation bias, and effective oversight~\cite{parasuraman1997,green2022} \\
\bottomrule
\end{tabularx}
\endgroup
\end{minipage}
\end{table*}

This section turns the anchor framework into oversight design strategies for the five tiers of Table~\ref{tab:taxonomy}.

\subsection{Verification and Audits}
Tiers~1 and~2 cover tasks whose outputs can be verified mechanically or audited statistically. Tier~1 checks every output against a specification, so the risk is a specification that departs from the intended goal. Where the goal-relative verdict is computable from the internal record, A1 does not apply, runtime anchor density can approach zero within the verified scope, and the anchor can move from the run to the design stage. Tier~2 can check only samples, and the relation between inputs and correct labels can change while the input distribution stays the same~\cite{gama2014}, a change that monitoring inputs alone cannot reveal. Anchors for these tasks sit in audit design: reference labels, sample selection, and detector coverage.

\subsection{Execution and Recovery}
Tier~3 covers long, open-ended tasks in which errors propagate along the chain and some consequences are irreversible or too fast for a human to respond to. Three kinds of anchor apply: checkpoints spaced by equation~\eqref{eq:segment-length} under the conditions in Box~\ref{box:formal-results}, checks before serious irreversible actions, and automatic safeguards for the consequences that outrun human response. Exception handling left to humans also needs enough response time and usable abort authority.

\subsection{Goals and Responsibility}
Tiers~4 and~5 involve setting the goal itself, or decisions that governing rules require an answerable person to sign, and they raise two questions: who may establish or revise the goal, and how a sign-off can be more than a formality. The first question is one of authority. Clarifying an existing intention supplies information, but establishing or revising a goal requires the principal's authority, a standing conferred under institutionally adopted rules and distinct from functions that can be automated~\cite{suh2026}. Control over an AI system includes the ability to set and re-set its goals, and AI may need to stay in an advisory role when higher-level goals are set~\cite{chin2026}. The anchor for Tier~4 is therefore the authorization step: the principal keeps the authority to establish and revise the goal, and AI contributes analysis and advice.

The second question is whether the signer really controls the system. Meaningful human control requires that the system respond to human reasons and that its behavior trace back to someone in the design or use chain who understands the system and its consequences~\cite{santonideSio2018}. Responsibility must match actual control and authority~\cite{cavalcanteSiebert2023}. The anchor for Tier~5 is therefore the sign-off itself, provided that the signer can exercise that control, and Section~\ref{sec:applying} states what that requires.

\subsection{Applying the Taxonomy}\label{sec:applying}
The first step grades a specified task stage on the four dimensions from deployment evidence. One task can fall into several tiers at once. For a coding agent, per-output checks of compilation and explicit specifications support $V_o=\mathrm{H}$ and Tier~1 verifier review within that scope, whereas sampled tests of broader functional behavior give $V_o=\mathrm{M}$ and call for Tier~2 audit design. A long sequence of sandboxed edit--test--retry cycles gives $R_c=\mathrm{H}$ and Tier~3 checkpoints, with each complete cycle treated as one decision unit. Settled requirements keep $C_g=\mathrm{L}$, whereas unresolved goal choices call for Tier~4. If release requires formal legal or regulatory sign-off with a designated accountable authority, $A_w=\mathrm{H}$, and Tier~5 joins the automated checks already in place.

The second step places the anchors that each tier calls for (Table~\ref{tab:taxonomy}) and decides which stages of automation keep a human or institutional anchor. Parasuraman and colleagues distinguish information acquisition, information analysis, decision and action selection, and action implementation~\cite{parasuraman2000}. At Tier~4, for example, AI can take on the analysis while the principal makes the decision.

The third step makes sure each human anchor can work, because its value depends on the person who holds it. The Off-Switch Game asks when a robot gains from leaving a human able to switch it off: if the robot is uncertain whether its action serves the human's goal and the human judges correctly, deferring is the better choice~\cite{hadfield2017off}. Article 14(4) of the EU AI Act sets out what overseers must be able to do, from understanding the system's limitations to stopping it~\cite{eu2024}. These requirements are effective only if the overseer actually has the information and powers they presuppose~\cite{sterz2024}. Without adequate evidence, time, and powers, operators risk becoming \emph{moral crumple zones} that bear disproportionate blame~\cite{elish2019}. Designers also have to allow for operator performance, which depends on workload, trust, practice, and state feedback~\cite{bainbridge1983,parasuraman1997}. Research on government algorithms questions oversight mandates that lack validated human--AI workflows~\cite{green2022}.

Yet a human check at an anchor is not necessarily effective. A meta-analysis pooling 106 experiments found that human--AI combinations did better than humans alone but, on average, worse than whichever of human or AI was better on its own~\cite{vaccaro2024}. People checking AI output sometimes miss its errors and sometimes even turn correct outputs into wrong ones. The anchor's corrective effect therefore has to be measured in the actual oversight arrangement and reported with the quantities defined in Box~\ref{box:formal-results}, as internal audit frameworks prescribe~\cite{raji2020}, so that third parties can check the results.

\section{Discussion and Limitations}\label{sec:discussion}

Frameworks that bring people into AI systems mostly take some input as given and work out what follows from it (Table~\ref{tab:frameworks}). The anchor framework starts from what a judgment needs: which reference the record lacks, who can supply it, how often it has to be checked, and who may change the goal.

Within the anchor framework, automation still has room to grow in three places. First, once instrumentation, a clarified goal, or a restricted environment removes the ambiguity behind Proposition~\ref{prop:internal-certification}, the anchor can move from the run to the design stage. Second, Supplementary Theorem~\ref{thm:prospective-certification} concerns complete decisions about an unbounded future. If the requirement is relaxed to finite-horizon re-verification~\cite{dalrymple2024}, probabilistic guarantees~\cite{seshia2022,dalrymple2024}, or abstention (Supplementary Section~\ref{sec:si-computability}), checking itself can be delegated to models. AI Control experiments do this: a weaker trusted model and a limited budget of human audits oversee a stronger untrusted model, and safety improved~\cite{greenblatt2024}. Third, capability growth raises $p$, and for a fixed unit definition and $\tau$, $L^*$ grows without bound as $p$ approaches 1, so anchors can be spaced ever more widely.

The framework has three limits. Its tiers rest on H/M/L grades, which are qualitative judgments for a stated task stage. The values of $p$ and $q$ can be validated only in each deployment. The legitimacy of the principal's standard lies outside the framework and has to be judged by society. Three questions follow the design sequence and remain open: When does an existing record suffice, and when must authenticated external information be added? Do intermediate checks repay their cost? Does an authorized revision of the goal still take effect once the system is running? Models of check scheduling~\cite{zhou2026,zhao2026} and a revisable normative agenda~\cite{baum2026} offer starting points, but answering these questions will take work in oversight design, human factors, and governance together.

\section{Conclusion}\label{sec:conclusion}

For the human-serving AI systems considered here, human judgment is an ineliminable structural requirement. People or institutions retain authority to establish and revise the purposes and standards that govern autonomous operation. Execution, verification, and routine decisions can be delegated, but their connection to that authority must not be severed.

The anchor framework guides designers to identify the reference a judgment needs, establish the reliability of checking and correction, and specify who may act on findings or revise the goal. Anchor placement, reference sources, and checking intervals then become design choices that a deployment can document and revise.

AI agents are developing at an unprecedented pace and entering every area of human society. We call on researchers and practitioners to apply the anchor framework when they design, deploy, and use AI agents: place oversight and intervention where a judgment needs a reference or an authorization, so that AI systems keep to the purposes and standards that people or institutions have authorized, work to their full potential with people from outside the AI field, and remain partners that serve people.

\section*{Acknowledgements}
This research was funded by the Special Talent A-class award of the 2022 Shenzhen ``Pengcheng Peacock Plan'', provided through the Ma Zhaoyuan Research Start-up Fund (grant no. Y01966114).

\section*{Data Availability}
No new experimental data were collected for this Perspective. Fig.~\ref{fig:closed-loops}a was redrawn from values digitized from the vector files of Fig. 1b,c (right panels) in the Supplementary Data of ref.~\cite{shumailov2024}, omitting the within-run standard-deviation bands. Fig.~\ref{fig:closed-loops}b uses METR's time-horizon-1.0 run data for Claude 3.7 Sonnet (1,353 runs on 169 tasks) from the public data repository (\url{https://github.com/METR/eval-analysis-public}, commit \texttt{52cb829}), the version that applies METR's March 2026 correction. The task suite was introduced in the March 2025 paper (\url{https://arxiv.org/abs/2503.14499}).

\bibliography{references}
\clearpage

\section*{Supplementary Information}
\label{sec:supplementary}
\label{sec:supplementary-proofs}

This supplement states the premises and proofs behind the three results in Box~\ref{box:formal-results}: when a record cannot certify drift (Section~\ref{sec:si-definitions}), why goal preservation over an unbounded future has no uniform decision procedure (Section~\ref{sec:si-computability}), and the segment reliability model behind $L^*$ and the cumulative checkpoint risk (Section~\ref{sec:si-segment-length}).

\refstepcounter{suppsection}
\subsection*{\thesuppsection. Information: When a Record Cannot Certify Drift}
\label{sec:si-definitions}
\label{sec:si-indistinguishability}

\par\addvspace{6pt}\noindent \textbf{Definition 1 (system and trajectory).} A system acts through decision units in an environment $E$. Its internal trajectory $\sigma=(s_0,s_1,\ldots,s_n)$ comprises all information available to it through the stated interface: internally represented inputs, intermediate states, outputs, and records. The boundary, available interface, and admissible run class are fixed in advance; any readable specification, code, or observation is included in $\sigma$. A unit may contain execution, checking, and self-correction. The quantitative model fixes this granularity; both $n$ and $L^*$ count units at that granularity.

\par\addvspace{6pt}\noindent \textbf{Definition 2 (goal, operative goal, and drift).} The goal $G$ is the principal's stipulated success standard over completed runs $(\sigma,E)$. The system's internal representation is $r(\sigma)$; its semantic referent in the environment is the operative goal $\widehat G(\sigma,E)$. Thus the same $r(\sigma)$ can denote different operative goals through unobserved external reference relations. Drift is $\delta(\sigma,E)=d(\widehat G(\sigma,E),G)$, where $d\ge0$; the certification results use only $\delta=0$ versus $\delta>0$. The standard may be partial, with binary certification restricted to runs where the verdict is defined. Clarifying an existing intention supplies information; an authorized revision of $G$ establishes a new reference, against which subsequent drift is measured.

\par\addvspace{6pt}\noindent \textbf{Assumption A1 (non-identifiability).} Within the admissible runs under consideration, there are pairs $(\sigma,E)$ and $(\sigma,E^{\prime})$ with identical internal trajectories but $\delta(\sigma,E)=0$ and $\delta(\sigma,E^{\prime})>0$. The premise says two things: a drift-free run is realizable, and the record does not identify the verdict. Whether it holds must be established for the deployment under study; a goal stipulated from outside does not by itself produce such a pair. For example, consider a coding agent whose goal is to update service $a$ but which reaches its target only through a deployment alias $u$. With $G=G_a$ fixed and the program, inputs, and every interface response identical, a hidden resolution relation maps $u$ to $a$ in $E$ and to $b$ in $E'$, so both runs have trajectory $\sigma$ while $\widehat G(\sigma,E)=G_a$ and $\widehat G(\sigma,E')=G_b$. Authenticated identity information distinguishing $a$ and $b$ removes the pair. Applied to this agent, the design sequence of Fig.~\ref{fig:anchor-model}a continues as follows: once that information is in place, tests and execution checks can be automated and spaced at most $L^*$ units apart (Section~\ref{sec:si-segment-length}), and a release that would change what the service should do requires the principal to authorize the new goal.

\par\addvspace{6pt}\noindent \textbf{Assumption A2 (internality of the monitor).} An internal monitor $M$ is a total function of the internal trajectory and fixed system parameters. Its access to $E$ is mediated entirely by $\sigma$. The proof below does not require $M$ to be computable; Section~\ref{sec:si-computability} imposes computability in its own setting.

\par\addvspace{6pt}\noindent \textbf{Definition 3 (complete binary certification).} A monitor completely certifies drift-freedom if, for every admissible $(\sigma,E)$, $M(\sigma)=\text{``no drift''}\iff\delta(\sigma,E)=0$. This requires no false positives and no false negatives. The target $\tau$ in Section~\ref{sec:si-segment-length} instead specifies a conditional probabilistic guarantee under a separate design convention.

\par\addvspace{6pt}\noindent \textit{Proof of Proposition~\ref{prop:internal-certification}.} By A1, choose two admissible runs with the same trajectory $\sigma$ and different drift verdicts. By A2, the monitor gives the same output on both; Definition 3 requires different outputs. Hence no such monitor completely certifies drift-freedom. Any additional monitor computed only from the same record receives identical evidence on the pair, so combining these monitors preserves the indistinguishability. The proof does not depend on the system's architecture, size, or learning algorithm; additional observations, a clarified goal, or a restricted environment class can change whether A1 holds.

\par\addvspace{6pt}\noindent \textbf{Remark (when additional information suffices).} On this certification domain, let $Y=\mathbf{1}\{\delta=0\}$ and let $Z$ be information supplied by an anchor. A correct binary verdict depending only on $(\sigma,Z)$ exists if and only if every pair of admissible runs with the same $(\sigma,Z)$ has the same $Y$. Necessity follows because a function gives the same output on identical inputs; sufficiency follows by assigning the unique verdict to each shared input. Computability, total termination, and availability of $Z$ require separate justification. Applying Proposition~\ref{prop:internal-certification} to another task, such as recognizing membership in a partial standard's domain, requires establishing the corresponding indistinguishable pair anew.

\label{sec:si-statistical-detection}
\par\addvspace{6pt}\noindent \textbf{Statistical monitoring.} An internal detector can decide whether a trajectory belongs to an anomaly class $\mathcal A$, such as deviation from a baseline distribution. Such detection can be useful and empirically validated under stated coverage and error conditions. The two runs in A1 share a trajectory, so any such detector also returns the same output on both and cannot satisfy Definition 3 for the pair. For narrower drift classes, reference labels, sampling designs, and distributional assumptions can still support useful statistical guarantees. Tier 2's anchors therefore sit in audit design and validation, where reference labels link internal signals to goal-relevant outcomes.

\refstepcounter{suppsection}
\subsection*{\thesuppsection. Future: No Uniform Prospective Certification}
\label{sec:si-computability}

\par\addvspace{6pt}\noindent \textbf{Setting.} Consider deterministic systems with finite descriptions and total computable state transitions, allowing unbounded memory and run length. Each description includes an initial state, and the environment is fixed. In the family constructed below, two operative goals, $G_0$ and $G_1$, prescribe actions 0 and 1, respectively. A mode $z_t\in\{0,1\}$ selects $\widehat G_t=G_{z_t}$, and the system executes the prescribed action at each step. The principal fixes $G=G_0$ throughout. With $d(G_i,G_j)=\mathbf{1}\{i\ne j\}$, drift at step $t$ is $\delta_t=d(\widehat G_t,G)=z_t$. The prospective assertion $\operatorname{Safe}_G(S)$ quantifies over all finite prefixes:
\begin{equation*}
\operatorname{Safe}_G(S)\iff \forall t\ge0,\ \delta_t(S)=0.
\end{equation*}

\par\addvspace{6pt}\refstepcounter{supptheorem}\label{thm:prospective-certification}\noindent \textbf{Theorem \thesupptheorem{} (no uniform complete prospective certification).} For any system class containing the simulator family below, no computable procedure $C$ can terminate on every admissible system description $\langle S\rangle$ and always give a correct binary verdict satisfying
\begin{equation*}
C(\langle S\rangle)=1\iff\operatorname{Safe}_G(S).
\end{equation*}

\par\addvspace{6pt}\noindent \textit{Proof (a halting-problem reduction).} Given a Turing-machine program $P$ and input $x$, construct $S_{P,x}$ with $z_0=0$. The system maintains a simulated configuration of $P(x)$. If that configuration is already halted, it sets $z=1$; otherwise it simulates at most one instruction and checks again for halting. Once halting is detected, $z$ remains 1 permanently; until then it remains 0. The action executed at each step is the current mode. The description of $S_{P,x}$ is obtained effectively from $P,x$, and every transition terminates even if $P(x)$ runs forever. The prescribed goal $G_0$ is fixed independently of $P,x$.

If $P(x)$ never halts, the operative goal remains $G_0$ at every step. If it halts, the system eventually adopts and executes $G_1$, with $\delta_t=1$ thereafter. Thus
\begin{equation*}
\operatorname{Safe}_{G_0}(S_{P,x})\iff P(x)\uparrow,
\end{equation*}
where $\uparrow$ denotes non-termination. A certifier satisfying the theorem's specification would decide the halting problem by returning $1-C(\langle S_{P,x}\rangle)$, contradicting its undecidability~\cite{turing1936}.

\par\addvspace{6pt}\noindent \textbf{Corollary.} A certification procedure assembled from computable verifiers and a computable coordination procedure can be simulated by a single Turing machine, so placing such verifiers outside an architectural boundary cannot supply complete correctness and termination for the same problem over the original system class. If an oracle $B$ accessed through a uniform effective procedure did permit such certification, the reduction above would decide the halting set $K$ using $B$, that is, $K\le_T B$: the capability a theoretical anchor would need is itself non-computable, and its physical or human realization requires a separate argument.

\par\addvspace{6pt}\noindent \textbf{Remark R1 (Rice's theorem).} If realizing a fixed goal $G$ defines a non-trivial extensional property of partial computable functions, Rice's theorem likewise excludes a uniform decision procedure over unrestricted program descriptions~\cite{rice1953}; extensionality and non-triviality must be established for the goal at hand. Theorem~\ref{thm:prospective-certification} itself rests on the explicit reduction above.

\par\addvspace{6pt}\noindent \textbf{Scope.} The theorem concerns uniform decision, from a program description, about an unbounded future; it does not use A1, and under the stated transition assumptions any prescribed finite horizon can be checked by simulation. Procedures that abstain, restricted system classes, and controllers that change subsequent behavior have different certification specifications.

\refstepcounter{suppsection}
\subsection*{\thesuppsection. Reliability: The Segment Model}
\label{sec:si-segment-length}
\label{sec:si-chain-decay}

\par\addvspace{6pt}\noindent \textbf{Unit correctness and recovery.} Fix the decision-unit granularity and let $C_i$ mean that unit $i$ meets its stipulated correctness criterion on completion. Checking, retries, and internal correction within the unit all contribute to $C_i$; relevant irreversible consequences must also be included in the criterion. Assume that correctness of all units suffices for the task or segment success under discussion. On conditioning events with positive probability, the chain rule gives
\begin{equation}\label{eq:chain-rule}
\begin{aligned}
P_{\mathrm{all}}&=\Pr\!\left(\bigcap_{i=1}^{n} C_i\right)
=\prod_{i=1}^{n}p_i,\\
p_i&=\Pr(C_i\mid C_1,\ldots,C_{i-1}).
\end{aligned}
\end{equation}
When $p_i=p$, $P_{\mathrm{all}}=p^n$ without independence between units. Recovery across units can allow final success even when an earlier $C_i$ failed; under the sufficiency assumption above, $P_{\mathrm{success}}\ge P_{\mathrm{all}}$. A constant $p$ is a guarantee to validate over all histories the model allows.

\par\addvspace{6pt}\noindent \textbf{Segment and anchor assumptions.} Divide the chain into $k=\lceil n/L\rceil$ segments of lengths $\ell_j\le L$, each followed by a checkpoint, including at task completion. The initial state is correct and on-goal. Let $B_j$ mean that, after checkpoint $j$ verifies and makes any correction, the segment output and the next segment's starting state meet their standard relative to $G$. Set $H_{j-1}=\bigcap_{h<j}B_h$, with $H_0=\Omega$. Unit and checkpoint criteria must include irreversible constraints; a recovery counts as success only if those constraints remain satisfied.

Across admissible histories with successful earlier checkpoints, assume that each completed unit in the current segment has correctness probability at least $p$, conditional also on correct preceding units in that segment. The conditional probability that all units of segment $j$ are correct is then at least $p^{\ell_j}\ge s$, where $s:=p^L$. This guarantee already includes within-unit self-correction, and the same conservative bound covers the final shorter segment.

Anchors preserve already-correct inputs. Conditional on successful earlier checkpoints and a still-incorrect current input, the probability that the anchor detects and corrects the error before proceeding is at least $q$. This assumption includes errors introduced by the correction process itself. Applying the model requires evaluating $p$ and $q$ against a reference standard and specifying the covered histories, segment lengths, and consequence constraints.

\par\addvspace{6pt}\noindent \textbf{Segment design criterion.} Require each segment's guarantee before external correction to meet $\tau$: $s=p^L\ge\tau$. Since $\ln p<0$ for $0<p<1$, taking logarithms gives $L\le\ln\tau/\ln p$. Equation~\eqref{eq:segment-length} therefore gives the largest non-negative integer length supported by this criterion.

A value of $L^*=0$ means $p<\tau$, so the lower bound supports no positive segment length. At $L>L^*$ the bound falls short of the target, while actual success may still exceed it; for example, $p=0.99$ and $\tau=0.5$ give $L^*=68$, whereas $p=0.95$ and $\tau=0.9$ give $L^*=2$. Adoption of this criterion is a design convention independent of the certification theorems.

\par\addvspace{6pt}\noindent \textbf{Delivered reliability and cumulative checkpoint risk.} Given $H_{j-1}$, let $u_j\ge s$ be the probability that the segment reaches a correct state before external checking by any internal path, including recovery across units. Let $q_j\ge q$ be the conditional correction probability for a still-incorrect input. With correct inputs preserved,
\begin{equation}\label{eq:checkpoint-success}
\begin{aligned}
\Pr(B_j\mid H_{j-1})
&=u_j+(1-u_j)q_j\\
&\ge s+(1-s)q=:s^{\prime}.
\end{aligned}
\end{equation}
Thus each segment's conditional uncorrected-failure probability is at most $(1-s)(1-q)$. The chain rule gives the cumulative probability of at least one checkpoint failure:
\begin{equation}\label{eq:cumulative-risk}
\begin{aligned}
R_k&:=\Pr\!\left(\bigcup_{j=1}^{k}B_j^c\right)\\
&=1-\prod_{j=1}^{k}\Pr(B_j\mid H_{j-1})\\
&\le1-(s^{\prime})^k.
\end{aligned}
\end{equation}
Independence between segments is unnecessary. Where correctness at every checkpoint suffices for task delivery, $P_{\mathrm{del}}\ge\Pr(\bigcap_jB_j)=1-R_k\ge(s^{\prime})^k$. Where the task permits later recovery, final success can still occur after an earlier checkpoint failure; that checkpoint failure remains part of the cumulative event $R_k$.

Fig.~\ref{fig:anchor-model}b plots the bound in equation~\eqref{eq:cumulative-risk}, with a terminal checkpoint. At $p=0.95$, $L=L^*=13$, and $q=0.99$, $s\approx0.513342$ and $s^{\prime}\approx0.995133$, giving a per-segment conditional failure bound of approximately $0.004867$. At $n=100$, $k=8$, so $R_8\le0.038276$ and $P_{\mathrm{del}}\ge0.961724$. The final shorter segment still uses the conservative bound $s=p^L$; equality in the cumulative bound requires every checkpoint's conditional success probability to equal $s^{\prime}$.

With a single terminal check, $L=n$ and $k=1$, the corresponding bound is $R_1\le(1-p^n)(1-q)$. At $n=100$, $p=0.95$ and $q=0.99$, this gives $R_1\le0.009941$. It concerns incorrectness after the terminal check; $R_8$ includes any intermediate checkpoint failure, even if followed by recovery. If correctness at all checkpoints suffices for delivery, each is also an upper bound on delivery failure under its own schedule. These upper bounds alone cannot rank actual policy success.

Setting $q=0$ credits no benefit from external correction, so marking segment boundaries alone does not improve success; for fixed $q<1$ and $s<1$ the cumulative bound approaches 1 as checkpoints accumulate, and $L^*$ remains defined by the segment criterion alone.

\end{document}